# MS-BioGraphs: Sequence Similarity Graph Datasets

Mohsen Koohi Esfahani    Paolo Boldi    Hans Vandierendonck    Peter Kilpatrick    Sebastiano Vigna

https://blogs.qub.ac.uk/DIPSA/MS-BioGraphs

*Abstract*—Progress in High-Performance Computing in general, and High-Performance Graph Processing in particular, is highly dependent on the availability of publicly-accessible, relevant, and realistic data sets.

To ensure continuation of this progress, we (i) investigate and optimize the process of generating large sequence similarity graphs as an HPC challenge and (ii) demonstrate this process in creating MS-BioGraphs, a new family of *publicly available real-world edge-weighted graph datasets with up to* 2.5 *trillion edges*, that is, 6.6 times greater than the largest graph published recently. The largest graph is created by matching (i.e., all-to-all similarity aligning) 1.7 billion protein sequences. The MS-BioGraphs family includes also seven subgraphs with different sizes and direction types.

We describe two main challenges we faced in generating large graph datasets and our solutions, that are, (i) optimizing data structures and algorithms for this multi-step process and (ii) WebGraph parallel compression technique. We present a comparative study of structural characteristics of MS-BioGraphs.

The datasets are available online on https://blogs.qub.ac.uk/DIPSA/MS-BioGraphs.

## I. Introduction

Because of the fast increase in the data production rate, and the existence of unstructured connections in these data, High-Performance Graph Processing (HPGP) has to date been widely applied in various fields of science, humanities, and technology. This fact has two main implications for the efficiency of public research and academia that aim to consider the real-world challenges and to design practically-applicable solutions to those challenges. The first effect is the necessity of having **realistic and up-to-date graph datasets** and the second implication is the necessity of considering the effects of new contributions (such as algorithms, processing models, parallelization, and data structures) on **a wide range of input datasets** to cover different application domains.

However, as we detail in Section II, the public graph datasets are small, domain-restricted, and not suitable indicators of real-world data which makes them not ideal for progressing HPGP.

To confront this problem, we investigate and optimize the HPC process of generating sequence similarity graphs and demonstrate this process in creating and introducing **MS-BioGraphs**, a new family of real-world graphs with up to 2.5 trillion edges that makes them **the largest real-world public graphs**. This family contains different graph sizes and direction types with similar structures that make them suitable for a range of applications with different input size requirements. Moreover, this graph family shows a very different graph structure in comparison to other real-world graphs (such as social networks and web graphs) and so, complements the current graph collection.

We faced two major challenges in optimizing (i) creation and (ii) compression of these large graphs. The creation of these large datasets is a multi-step process in which (a) the dependency between steps and (b) the processing requirements (i.e., availability of processing resources, memory, and storage) should be considered in the selection and creation of data structures and algorithms of each step. The flow of data between different steps of a multi-step process have important effects on the processing efficiency of the steps. As such, the whole process and processing requirements should be considered and be optimized by **process-wide engineering and design of data structures and algorithms**.

The processing model is one of the main choices in this optimization. The distributed-memory processing model [1], [2] implies two restrictions: (i) fixing the degree of parallelism (i.e., the number of machines/processors involved in the processing) and (ii) limiting the size of processed data to the total memory of the cluster. On the other hand, the storage-based processing model [3], [4] does not practically limit the size of data but deploys only one machine and increases the processing time. Therefore, we designed the processes as multi-step tasks where each step is performed as a distributed parallel computation but without communication between machines. Machines process the partitions independently from each other and use the cluster's shared storage for loading and storing the (intermediary) data.

The second major problem is efficient compression of graph datasets to facilitate fast transfer of the created datasets. The WebGraph Framework [5] provides graph compression at high scale, but the compression process is sequential and we **extend the WebGraph framework by parallelizing compression**.

We study some **features of the MS-BioGraphs** showing that (i) while these large biographs follow a skewed degree distribution (similar to other real-world graphs), they expose a different arrangement of edges in comparison to previous graph types by having tight connections between the frequently-occurring high-degree vertices that make their graph structure distinct from other real-world graph types, (ii) weights have a skewed distribution with a tail close to power-law distribution, (iii) the main graph and its large subgraphs exhibit a high-degree of connectivity, and (iv) the asymmetric MS-BioGraphs have a close Push and Pull Locality which is different from social networks and web graphs.

The contributions of this paper are introduction of:
- the HPC-optimized multi-step process of creating large sequence similarity graphs,
- the MS-BioGraphs family as the largest real-world public graphs and publishing them as open datasets,
- parallel compression in the WebGraph framework, and
- a structural analysis of MS-BioGraphs.

This paper is structured as: Section II motivates the discussion by exploring the needs for large real-world graphs and considering their effects on progressing HPGP. Section III introduces the processing model and parallel graph compression as our solutions for the major challenges in processing large graphs. Section IV explains the creation process of large graphs and demonstrates it for creating MS-BioGraphs. Section V presents a structural analysis of MS-BioGraphs and compares them with other types of real-world graphs. Section VI discusses related work and Section VII concludes the paper and expresses future directions.

## II. MOTIVATION

In this section, we consider (i) the necessity of creating updated and cross-domain datasets, (ii) the impacts of these datasets on the progress of HPGP, and (iii) the features of an ideal graph dataset.

### A. Why Do We Need Updated and Real-World Graphs?

(1) While synthetic graph generators [6], [7] can create large graphs, *the structural features of synthetic graphs do not match the real-world ones*. E.g., they may expose several gaps in the degree distribution [8] and randomly selected vertices have a large percentage of similar neighbors. As such, the severity of challenges relating to partitioning, locality and load balance in synthetic graphs is often much lower than in real-world datasets. Therefore, the techniques that are sufficient for synthetic graphs may not be applicable for real-world datasets.

(2) Some graph optimizations are dependent on the architecture of machines and it is the tension between data size and the architecture capacities that forms the challenge context and presents the opportunity to design novel data structures, algorithms and processing models. E.g., the design of locality-optimizing algorithms [9], [10], [11], [12] depends on the fact that CPU's cache contains a small portion of the data. By the advent of CPUs with cache sizes of multiple GigaBytes, the locality optimizing algorithms play no role for small datasets as accesses to a large portion of data is covered by cache. Similarly, the progress of distributed graph processing [1], [2] may be slowed down by increase in per-machine memory capacity that is enough to host available datasets. This shows that *without large real-world datasets, it is not possible to progress these architecture-competing HPGP activities*.

(3) Several HPC research fields (such as architecture design, distributed and disk-based processing, and high-performance IO) have tight connections and dependencies on graph algorithms and datasets. *The effectiveness and realness of graph datasets guarantees the efforts on the dependent fields to have real-world impacts*.

(4) Creating a real-world graph dataset provides a representation of the data that *acts as a new source for extracting domain-specific information and knowledge* by deploying graph algorithms. As an example, sequence similarity graphs have several usages in biology including sequence clustering [13], [14], [15], [16], predicting pseudogene functions [17], effective selection of conotoxins [18], predicting evolution [19] and gene transfer [20].

A comprehensive graph representation of the data is also beneficial (i) to validate previous hypotheses (that have been verified on a small portion of data) in a wider perspective and (ii) to provide new opportunities to make new contributions by considering the new patterns and connections revealed in graph representation.

### B. Why Do We Need Different Types of Real-World Graphs?

(1) Previous studies have shown that different real-world graph types exhibit contrasting behaviors with graph analytic algorithms and optimizations [12]. Examples include the long execution time of small road networks in Label Propagation Connected Components [21], [22], the different impact of similarity and locality in web graphs and social networks [23]. This implies that a wider range of graph types will be necessary *to better study and comprehend the structure of graphs and to compare them*. This better understanding of different graph types and their structures will also be helpful to *design synthetic graph generators with greater similarity to real-world graphs* (Section II-A).

(2) A wide range of real-world datasets facilitates *cross-domain evaluation of the new contributions* and provides broad and correct assessment across a variety of use cases (i.e., better pruning of the falsifiable insights [24]). Also, we will have the opportunity to improve several graph algorithms and optimizations that exploit the structure of graphs [2], [10], [11], [25], [26].

### C. Creating Real-World Graphs: An HPC Problem

(1) Creating real-world graphs is a time-consuming process [27], [28], [29] and is periodically repeated. As the size of input dataset (connections in web graphs, links in social networks, or similarities in sequences) grows, greater amounts of computations and processing resources are required.

(2) Some tasks in creating graphs are widely used in deploying graph algorithms, such as format conversion, transposition, and symmetrization, are time-consuming [30]. Optimizing these steps is directly transferred in graph algorithms.

*D. The Current Largest Graph Datasets*

At present, the last largest public graph dataset we are aware of is the Software Heritage 2022-04 version-control-history graph[1] [28] with 376 billion edges that was published in 2022.

The largest web graph is Web Data Commons 2012 hyperlink graph[2] [29], with 128 billion edges that was published about 9 years ago. The largest social network graph is a snapshot of Twitter on 2010 [31] with 1.5 billion edges.

These graphs are outdated and/or not an indicative of the growth in size of data that is happening in the real world.

*E. What Is An Ideal Graph Dataset?*

The discussions in this section show that a new family of graphs should ideally (i) be backed by a real-world phenomenon, (ii) cover a wide range of graph sizes to make it suitable for different applications, (iii) exhibit new structural features that are not seen in other real-world graphs, (iv) contain graphs much larger than existing ones and in line with the exponential growing rate of the worldwide datasets [3], and (v) be available as open datasets to research communities.

## III. HPC CHALLENGES AND OUR SOLUTIONS

In this section, we present two major challenges we faced in creating large datasets. Section III-A explores how to efficiently utilize a small cluster for processing large datasets. Section III-B explores how to parallelize the compression process of the large weighted graph datasets. We demonstrate our solutions for these two challenges in Section IV where we detail creating MS-BioGraphs.

*A. The Processing Model*

We search for a processing model that *(i) dynamically adjusts the degree of parallelism (i.e., the number of machines/processors involved in the processing) and (ii) does not restrict the size of processed data to the total memory of the cluster* while machines have access to a shared storage that hosts the datasets and the intermediary data.

The distributed-memory processing model [1], [2] sets an upper bound for the size of dataset based on the total memory of the cluster. This model also makes the waiting time of jobs dependent on the size of the requested resources. If we need a greater number of machines, we may need to wait for a longer time before scheduling the job. Therefore, to optimize cluster utilization it is necessary *to minimize the waiting time*.

The storage-based processing model [3], [4], on the other hand, does not practically limit the size of data, but deploys one machine and increases the processing time.

To satisfy the mentioned requirements, we deploy a distributed model in which algorithms are designed as a number of sequential steps with parallel workloads per step. In each step, machines contribute to the total processing independently from each other and the input and output data for each processing slot is loaded from and stored to the shared storage. So, machines only communicate (a) to the shared storage to retrieve/store data and (b) to the scheduler to receive a partition of a task or to inform completion of a partition.

In this way, each machine requires a memory size that is enough to complete a partition. This facilitates processing the datasets whose sizes are greater than the available memory.

Moreover, as the machines do not communicate with each other, each step can be started as soon as at least one machine becomes available and new machines can join/leave a running step. This (i) relaxes the assumption of permanent availability of a fixed number of resources during the whole execution time, (ii) minimizes the waiting time, and (iii) optimizes cluster utilization.

*B. Parallelizing Graph Compression*

As MS-BioGraphs have binary sizes of up to 20 TeraBytes, it is necessary to compress them to make their storage, transfer over the network, and processing more efficient.

To that end, we used the WebGraph framework[4] [5] which is an open-source graph compression framework that has been continuously maintained and updated during the last 20 years. This framework provides a graph compression and includes a rich set of graph operations and analytics. Moreover, the users of languages and frameworks with WebGraph support, such as Hadoop, C++, Python, and Matlab, benefit from direct access to MS-BioGraphs.

WebGraph provides facilities for storing edge-labelled graphs. Labels are stored contiguously in a bitstream in edge order (i.e., lexicographical source/destination order), and an offset file containing pointers to the start of the sequence of labels associated with the neighbors of a vertex. The bitstream can be loaded into memory or memory-mapped to support graphs with a larger size than core memory. Moreover, offsets are loaded using the Elias–Fano representation, a quasi-succinct data structure that brings the required storage space for each offset to a few bits [32].

Historically, the design of the labelled facilities in WebGraph decoupled the compression of the underlying graph and the storage of the labels. This approach has the advantage of implementing a clear separation of concerns and makes it possible to pair compression schemes and label storage schemes arbitrarily.

However, in processing MS-BioGraphs, it became clear that the approach is very inefficient in a number of situations, and in particular when transposing, symmetrizing or permuting very large labelled graphs. In all of these operations, graph edges are first divided into batches that are sorted in core memory using a parallel sorting algorithm and compressed on disk; then, one can traverse the resulting transposed (or symmetrized, or permuted) graph sequentially. However, this traversal is quite expensive as the compressed representation is optimized for space and ease of storage, but not for speed of traversal; ideally, the transformed temporary graph should be traversed exactly once.

---

[1]https://docs.softwareheritage.org/devel/swh-dataset/graph/dataset.html
[2]http://webdatacommons.org/hyperlinkgraph/2012-08/download.html
[3]https://www.idc.com/getdoc.jsp?containerId=US49018922 and https://www.statista.com/statistics/871513/worldwide-data-created

[4]https://webgraph.di.unimi.it/

The previous design was thus at odds with this approach, as two passes were necessary to compress the graph and to store the labels. Moreover, the current implementation of labelled graphs did not allow for parallel storage—a fundamental requirement in processing large-scale graphs.

We extended the WebGraph framework in two directions: in the first phase, we extended labelled graphs to support parallel compression of the underlying graph. This first extension decreased significantly the compression time (scaling is linear in the number of cores) but did not solve the problem of multiple passes over the temporary representation.

In the second phase, we partially violated the decoupled design of labelled graphs in WebGraph, adding to the compression phase of the main storage format class of WebGraph, `BVGraph` (that compresses and stores the underlying graph), an option to store the labels at the same time. This created a dependency of `BVGraph` on a *specific* labelled graph implementation; that is, the parallel and simultaneous compression of graph and labels can only happen with a specific, bitstream-based label representation. However, since recompressing the underlying graph in a different format can be performed with very low cost, and the bitstream-based label implementation is the only presently-available option, the implementation remains, in practice, highly (albeit not completely) decoupled.

## IV. GENERATING MS-BIOGRAPHS

In Section III, we introduced solutions for major challenges in processing large graphs. In this section, we demonstrate those solutions to design and implement the algorithms required in different steps of creating the MS-BioGraphs.

### A. Terminology

A directed graph $G = (V, E)$ is defined by a set of vertices $V$ and a set of edges (a.k.a. arcs) $E \subseteq V \times V$; an edge is an ordered pair $(u, v)$ that indicates an edge from vertex $u$ to $v$. In a directed weighted (a.k.a., labelled) graph $G_w = (V, E)$, the set of edges is a subset of $V \times V \times \mathbb{N}$, where $(u, v, w) \in E$ represents an edge from $u$ to $v$ with weight $w$. The *undirected weighted graph* $G_u = (V, E)$ is defined as a directed weighted graph where for each $(u, v, w) \in E$, there is an edge $(v, u, w) \in E$.

A protein sequence is a string of letters, each letter representing one of the 20 canonical amino acids. Each of these 20 amino acids is represented by a letter from "ACDE-FGHIKLMNPQRSTVWY"[5]. A *sequence similarity matching* or a sequence aligning algorithm is an algorithm that receives two sequences as inputs and outputs a number that represents the similarity of the input sequences.

Similarity is calculated by comparing aligned amino acids whose matches are not directional and match values are derived from a symmetrical matrix (e.g., PAM and BLOSUM). Therefore, similarity using standard approaches (e.g., Smith-Waterman) is undirected, i.e., $Similarity(S_1, S_2) == Similarity(S_2, S_1)$. Also, two sequences may have multiple matches as the start point of match is not restricted.

TABLE I: Machines

|  | **SkyLakeX** | **SkyLakeX-2** | **Haswell** | **Epyc** |
|---|---|---|---|---|
| CPU Model | Intel Xeon Gold 6130 | Intel Xeon Gold 6126 | Intel Xeon E5-4627 | AMD Epyc 7702 |
| CPU Freq. (GHz) | 2.10 - 3.7 | 2.6 - 3.7 | 1.2 - 2.6 | 2.0 - 3.35 |
| CPU Cores/Machine | 32 | 24 | 40 | 128 |
| Memory/Machine | 768 GB | 1,536 GB | 1,024 GB | 2,048 GB |
| Number of Machines | 1 | 2 | 2 | 4 |

For a set of protein sequences, the *sequence similarity graph* is a weighted undirected graph whose vertices represent proteins and with an edge $(u, v, w)$ expressing the fact that the similarity between proteins $u$ and $v$ (the endpoints of the edge) is $w$; in other words, the weight of an edge is the similarity score calculated by the sequence aligner algorithm. It is important to note that the sequence similarity graph is *not a clique*, as only edges with a minimum level of similarity are added to the graph (the aligner algorithms only produces an output if the two sequences can be matched).

### B. Input Dataset & Environment Setup

Inspired by HipMCL [13], we use the Metaclust dataset[6] [33] that contains 1.7 billion protein sequences in FASTA format[7]. We collected all similarities produced by the LAST sequence alignment algorithm[8] [34] Version 1293. We selected LAST as aligner as it shows better single-machine performance [35] and has been widely used and maintained since its publication in 2011.

Sequence matching by LAST is performed in two steps: (i) creating a database (DB) from sequences using a program called `lastdb` and (ii) aligning the sequences of a file against the created database using `lastal` (with `PAM30` scoring matrix and default values for other options) that outputs the matched sequences and their scores.

Table I shows the hardware used in this project for all experiments; they all have CentOS 7.9 installed. Since the mentioned computers are setup in a job-sharing cluster, not all machines (and not all of their cores and memory capacity) were permanently available in all steps. So, for each step we report the machines that were used and their processing times. The cluster is backed by a 2 PetaBytes Lustre file system that provided up to 8 Gbps bandwidth in our experiments.

We have implemented most of our algorithms as extensions to the LaganLighter framework[9] [36], in the C language with `OpenMP` parallelization. We also used the `libnuma` library with the interleaved NUMA memory policy. Our algorithms were compiled with `gcc-9.2` using the optimization `-O3` flag.

---

[5] https://en.wikipedia.org/wiki/Amino_acid
[6] https://metaclust.mmseqs.com/2018_06/metaclust_all.gz
[7] https://en.wikipedia.org/wiki/FASTA_format
[8] https://gitlab.com/mcfrith/last
[9] https://blogs.qub.ac.uk/DIPSA/LaganLighter/

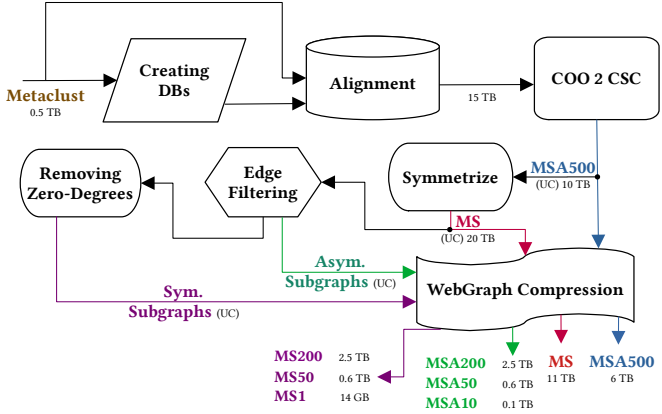

Fig. 1: Creation Steps (UC: uncompressed)

The parts of our algorithms that exploit WebGraph required to run Java, and we used `JDK-17` and `OpenJDK-19-Loom` that provides the `incubator.foreign` package to facilitate frequent file mapping using the `MemorySegment` class.

Our processing model requires a dispatcher to track the steps and to assign partitions of job in each step to machines. The schedulers that support job dependencies (such as Slurm and OpenPBS) can be used as this scheduler. We have implemented the dispatcher as a PHP script that is backed by an Apache server and a MySQL database.

## C. Process-Wide Data Structures and Algorithms Engineering and Design

In this section, we design the general process of creating MS-BioGraphs as a multi-step process by considering the flow of data and dependencies of steps. We explain the detailed algorithms and implementations of each step in Section IV-D.

(1) To create MS-BioGraphs, we compute all-against-all matching of the sequences. Since sequence similarity is a symmetric relation, instead of matching each pair of sequences twice, we can match each sequence only to sequences with lower IDs. This produces a directed weighted graph whose symmetric version represents all the matches and their scores. This imposes the cost of symmetrization but reduces the alignment computations by 50%.

We have the following steps as depicted also in Figure 1. First, we need to create LAST database(s) using `lastdb` and then call `lastal` to create the similarities, i.e., the asymmetric graph in the coordinate format (COO). The next step is converting the COO graph to the Compressed Sparse Columns (CSC) [37] format which is followed by symmetrizing and compression. We also create some subgraphs to support research studies with different graph size and direction requirements. Therefore an "Edge Filtering" step is required to create subgraphs and we need to remove zero-degree vertices.

(2) We need to consider whether to run the `lastal` in parallel mode on one single machine. Our preliminary evaluation showed that the `lastal` does not continuously engage all processors. The other problem is the long processing time (366 hours as we report in Section IV-D) as a result of deploying one machine.

However, there is a more important implication of running one instance of `lastal` and that is its output. The output of the "Alignment" step is used as input to the "COO to CSC" step. The CSC format consists of two arrays: the `offsets` array and the `edges` array. The `offsets` array is indexed by a vertex ID to identify the index of the first edge of that vertex in the `edges` array. In creating the `edges` array, we need to read edges from the COO graph and to write each edge based on the offset identified by its destination endpoint. This requires random write accesses to the `edges` array which requires 8 Bytes per edge (4 bytes for the ID of the source endpoint and 4 Bytes for the weight), or about 10 TB memory.

As no machine has this size of memory, the other option is to convert the subgraphs of the COO format to the CSC subgraphs and then merge the CSC subgraphs to create the CSC graph. While this can be done in a distributed way (Section III-A), it implies one extra reading and one extra writing of all edges.

So, we face three problems: (i) load imbalance of `lastal` in parallel mode, (ii) long execution time in the "Alignment" step, and (iii) storage overhead in the "COO to CSC" step.

Our solution for this cross-step problem is to partition the input dataset that converts the adjacency matrix of the graph to a number of blocks. The graph construction is now performed by calling concurrent instances of `lastal` for different blocks, (i.e., pair of partitions) and each instance is run in sequential mode. This optimizes load balance, increases the cluster utilization, and significantly reduces the computation time by concurrently deploying multiple machines (Section III-A).

Each block of the adjacency matrix is stored in a separate file and allows us to efficiently create the CSC graph in the distributed model by partially creating the CSC graph for each partition where it is only needed to load the relevant blocks (for partition $p_j$, all edges exists in $(p_i, p_j)$ blocks where $i \leq j$) and we do not need to keep the whole `edges` array in the memory. By having a sufficiently large number of partitions, we ensure the memory space required for a slice of the `edges` array is available on each machine.

(3) The output of "COO 2 CSC" is symmetrized to create the main graph. This is efficiently done in the distributed model by transposing and merging the transposed graph with the CSC graph. It is possible to merge the "COO 2 CSC" and "Symmetrize" steps into one step by transposing each partition while creating the CSC format and then merging the transposed subgraphs and CSC. However, this results in concurrency of two write and one read storage operations for all edges that may overload the storage bandwidth. Our evaluation shows that overloading storage bandwidth in our cluster (with per-user bandwidth limit) imposes longer delays. However, merging these steps is beneficial for clusters that provide greater storage bandwidth limit.

(4) "Edge Filtering" and "Removing Zero-Degrees" are efficiently done in the distributed model. The last step is creating

compressed version in WebGraph format which deploys a shared-memory model.

### D. Processing Steps

**Step 1: Assigning IDs to Sequences & Creating DBs.**

We divide the input dataset (of size 471 GB), which is in the name-sequence format, into 120 ID-sequence partitions by replacing the name of each sequence with its ID so that we can use the output of `lastal` without converting the names of sequences to their IDs. Then, `lastdb` is called for each of these 120 ID-sequence files. We write the ID and name of sequences of each partition in an ID-name file so that the results of analytics on the sequence similarity graphs can be used to identify the name of sequences using their IDs. We used a shared-memory parallel implementation for this step that runs on one Epyc machine. Then, all instances of `lastdb` are called concurrently to create the databases of 120 ID-sequence files. While multiple machines could be used for processing required in `lastdb`, the small number of partitions (i.e., 120) made it enough to deploy one machine.

**Step 2: Sequence Aligning.** We align sequences of each partition against partitions with smaller IDs. This involves running `lastal` for 7 260 pairs of partitions.

We launch `lastal` in the distributed model (Section III-A) by implementing a CPU-load meter program that continuously compares the load of allocated processors. If the processors are not completely busy, a job (i.e., a pair of partitions) is requested from the dispatcher and is passed to a `lastal` instance. We slightly modified `lastal` in order to (i) receive two additional arguments that indicate the two partitions that are matched, and (ii) to have an additional output that, for each two matched sequences, writes their IDs and the resulting score (in binary format) in a file. The binary output files are named by the ID of aligning partitions (i.e., the additional input arguments). In this way, for each pair of partitions, say $(p_i, p_j)$ where $p_i \leq p_j$, the output files contain a set of tuples. Each tuple $(x, y, z)$ indicates that $x$ and $y$ are the sequence IDs ($x \leq y$) and z is their matching score. This creates the COO graph in a collection of files, each named based on the ID of partitions. The total COO files is 15 TB.

This step was completed by using 8 machines (3 Epyc machines and 5 Intel ones). The total processing is 366.3 machine-hours or 45.7 hours per machine, on average.

**Step 3: Converting COO to CSC.**

The CSC format consists of (i) the $offsets$ array which is the prefix sum of the in-degrees and (ii) the $edges$ array. The $offsets$ array identifies which section of the $edges$ array holds the in-neighbors of a vertex.

To convert the COO graph (stored in multiple files) to CSC format, three phases are required: (i) performing one pass over the COO files and counting the degree of destinations, (ii) calculating the prefix-sum of the degree of vertices to specify the `offsets` array (which is written to the secondary storage in order to protect from the following changes), and (iii) a second pass over the COO files that writes each edge in the `edges` array in the index obtained from the `offsets` array (indexed by the ID of the destination vertex to get the insertion point, which is then atomically incremented) and sorting neighbors of each vertex (based on their IDs) before writing to the secondary storage.

The second pass involves random write accesses to the `edges` array. However, by the special arrangement of the COO files (explained in Section IV-C) it suffices to have a memory space that is large enough to host edges of the partition(s) that are processed together and it is not required to host the whole `edges` array.

In the first pass, the COO files of each partition are read to identify the degree of vertices. In the second pass, we grouped the 120 partitions of the vertices into a number of groups, where vertices in each group require about 1 TB memory space for their edges. We load the COO files of the partitions in each group, after processing them in memory, we write the processed edges to the relevant offsets of the `edges` file.

As the performance of the algorithm is only dependent on the storage and as parallel threads of one machine are sufficient to saturate the bandwidth of the shared storage in our cluster, we implemented a shared-memory parallel model. However, it is easily integrated into the distributed model (Section III-A) by applying a modification to retrieve the partition group ID from the dispatcher instead of processing all groups sequentially. We used one Epyc machine for the parallel processing of this step that completed in 20.4 hours.

To confirm the correctness of the CSC graph, we designed and implemented a validation algorithm in the distributed model. Each machine requests a partition from the dispatcher and loads the CSC edges of vertices in that partition to the memory. Then, the COO files of this partition are read by parallel threads and, for each edge in a COO file, a binary search is performed between the edges of the destination vertex in the CSC format (which has been partially loaded as the `edges` array of the CSC graph). The validation completed in 18.6 machine-hours using 4 Epyc machines.

**Step 4: CSC to Compressed WebGraph Format.**

In this step, we convert the binary CSC graph to the compressed WebGraph format. We implemented an extension of the `ArcLabelledImmutableGraph`[10] class with random accesses to the edges in order to parallelize the compression (Section III-B). We used one Haswell machine and the task completed in 19 hours.

It is necessary to mention that two sequences can be matched by `lastal` with two or more scores. Therefore, the graphs created in Steps 2 and 3 have some edges with the same endpoints but different weights. As these *same-endpoints edges* were less than 1% of the total edges[11], we selected the weight with the largest value for these edges and the compressed WebGraph format has at most one edge between each ordered pair of vertices.

---

[10]https://webgraph.di.unimi.it/docs/it/unimi/dsi/webgraph/labelling/ArcLabelledImmutableGraph.html

[11]Exactly 5,035,492,026.

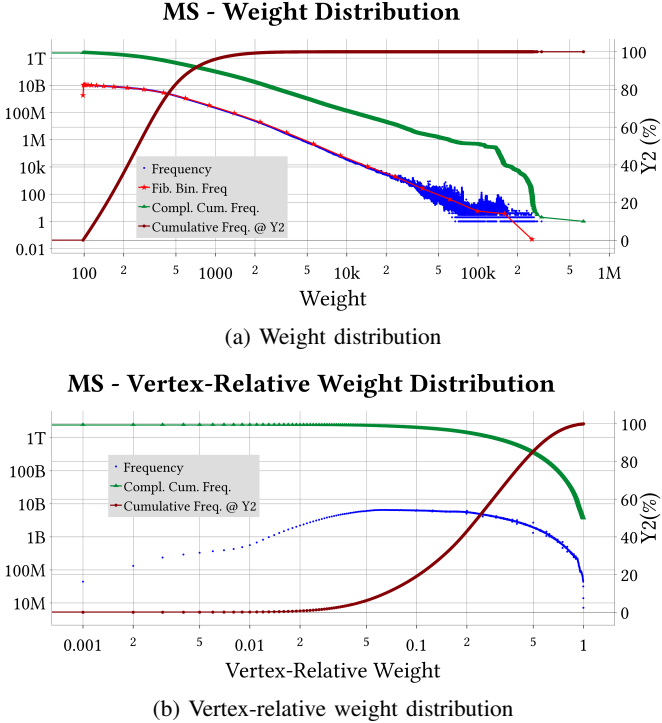

(a) Weight distribution

(b) Vertex-relative weight distribution

Fig. 2: MS graph weight distributions

As we explain in Section V, this directed graph (that contains only similarities to neighbours with lower or equal IDs) is called **MSA500** and its symmetric version (that is created in the next step) is called **MS**.

**Step 5: Symmetrizing.**

To create the MS graph, we compute the symmetric version of the MSA500 in two ways: (1) by using the WebGraph framework and (2) by designing and implementing a distributed algorithm (Section III-A) to compute the symmetric graph in binary CSC format. The distributed algorithm works in three steps: (i) it divides the vertices of the MSA500 graph into partitions (i.e., subgraphs), transposing each partition, and storing the transposed versions on secondary storage in the binary CSC format, (ii) it creates the offsets array of the symmetric graph by calculating the degree of each vertex (that is, the sum of its degrees in the asymmetric graph and in transposed subgraphs) and computing the prefix sum, and (iii) it creates the edges array of the symmetric graph that for each vertex of a partition includes edges from the asymmetric graph (MSA500) and the transposed subgraphs (the start index for the edges of a vertex is identified by the calculated offsets array). Edges are sorted either on the first step (which is more work-efficient) or on the last step. The algorithm ran on 4 Epyc machines and the total processing was 160 machine-hours. We also validated the two symmetric versions (the WebGraph format and the binary CSC format) by matching the degree and edges of all vertices.

**Step 6: Edge Filtering & Removing Zero-Degrees.**

By the end of Step 5, we have the MS graph with 2.5 trillion edges and the MSA500 graph with 1.2 trillion edges.

To support a larger extent of users with varying processing models/needs and storage/memory limits, we decided to create smaller subgraphs by filtering edges.

To create **undirected subgraphs**, we used the cumulative weight degree distribution of the MS graph (Figure 2a) to identify 3 weight borders in order to create subgraphs with 20%, 5%, and 0.1% of the total edges that are called **MS200**, **MS50**, and **MS1**, respectively.

As removing edges by considering weights may remove all edges of vertices that do not have enough large weights, we considered another sampling method by considering **vertex-relative weights**. In this method, for each vertex we identify the maximum weight and then remove the edges whose weights are smaller than a fraction of the maximum weight of the vertex. As a result, an edge $(u, v, w)$ may be removed when considering it as an edge for $v$, but its symmetric version $(v, u, w)$ may remain after filtering as an edge of $u$ that has a lower maximum weight.

As a result of considering the vertex-relative weights, **directed subgraphs** are created. We used the vertex-relative weight degree distribution of the MS graph (Figure 2b) to identify three borders to create directed subgraphs with 20%, 5%, and 1% of the total edges that are, respectively, called **MSA200**, **MSA50**, and **MSA10**.

To create these 6 subgraphs, we designed a distributed algorithm which is similar to the symmetrizing step. First, the graph is divided into partitions and for each partition, edges are traversed, filtered, and stored on secondary storage. So for each partition, 6 sub-partitions (3 directed and 3 undirected ones) are created. In the second step, for each 6 target subgraphs, the related stored sub-partitions are merged to create the subgraph. In this way, by making one pass over the edges of the MS graph, all 6 subgraphs are created.

As a result of weight-based filtering in creating the undirected subgraphs, the zero degree vertices increased to 19%–97%. To **remove the zero degree vertices**, we designed a shared-memory parallel algorithm that first identifies the zero degree vertices and creates the vertex-renumbering array[12]. By removing the zero-degree vertices from the offsets array, the new offsets array is created. The new edges array is created by assigning the new neighbour IDs using the renumbering array. We used 4 Epyc machines for filtering step and the total processing required 31.1 machine-hours. The validation also finished in 27.6 machine-hours. The execution of zero-degree removal step for three undirected subgraphs (MS200, MS50, and MS1) using one Epyc machine completed in 2.4 hours. The validation process completed in 2.3 hours on one Epyc machine.

---

[12] The renumbering array is indexed by the old ID of a vertex and returns its new ID (in the graph with removed zero-degrees). We publish the reverse array (new-ID to old-ID) so that names of vertices (sequences) can be identified.

TABLE II: MS-BioGraphs Statistics - Abbr.: Kilo (**K**), Million (**M**), Billion (**B**), and GigaBytes(**GB**) - **W** and **VRW** in Column 5 indicate Weight and Vertex-Relative Weight - **Avg. Deg.** indicates Average Degree - Column **Weak. Con. Comp.** shows the number of Weakly-Connected Components and the relative size of the largest component - Column **Size** shows the size in secondary storage for the base (underlying) and labels (weights) graphs in WebGraph format.

| Name | Directed | \|V\| (M) | \|E\| (B) | Filtering Intention | Max. Deg. | | Weight | | Zero Deg. | | Avg. Deg. | | Weak. Con. Comp. | | Size (GB) | |
|---|---|---|---|---|---|---|---|---|---|---|---|---|---|---|---|---|
| | | | | | In(K) | Out(K) | Min. | Max. | In(M) | Out(M) | In | Out | Count(M) | Max. Size(%) | Base | Labels |
| **MS** | No | 1,757.3 | 2,488.0 | - | | 814.9 | 98 | 634,925 | | 6.4 | | 1,415.8 | 148.9 | 99.95 | 6,843.6 | 4,696.0 |
| **MS200** | No | 1,414.4 | 502.9 | 0.200\|E\|, W | | 745.7 | 460 | 634,925 | | 0.0 | | 355.6 | 338.3 | 96.61 | 1,362.7 | 1,119.6 |
| **MS50** | No | 585.6 | 124.7 | 0.050\|E\|, W | | 507.8 | 900 | 634,925 | | 0.0 | | 213.1 | 155.3 | 81.95 | 327.1 | 303.1 |
| **MS1** | No | 43.1 | 2.6 | 0.001\|E\|, W | | 14.2 | 3,680 | 634,925 | | 0.0 | | 61.7 | 15.7 | 4.66 | 6.1 | 7.7 |
| **MSA500** | Yes | 1,757.3 | 1,244.9 | ID$_{neigh}$ ≤ ID$_v$ | 229.4 | 814.4 | 98 | 634,925 | 6.4 | 16.8 | 711.0 | 715.3 | 148.9 | 99.94 | 3,502.2 | 2,351.8 |
| **MSA200** | Yes | 1,757.3 | 500.4 | 0.200\|E\|, VRW | 658.8 | 709.1 | 98 | 634,925 | 6.4 | 7.4 | 285.8 | 286.0 | 221.5 | 99.29 | 1,455.2 | 1,033.7 |
| **MSA50** | Yes | 1,757.3 | 125.3 | 0.050\|E\|, VRW | 543.1 | 297.9 | 98 | 634,925 | 6.4 | 8.5 | 71.6 | 71.7 | 363.1 | 94.15 | 385.2 | 268.3 |
| **MSA10** | Yes | 1,757.3 | 25.2 | 0.010\|E\|, VRW | 207.2 | 62.0 | 98 | 634,925 | 6.4 | 9.9 | 14.4 | 14.4 | 628.5 | 61.72 | 84.0 | 57.3 |

## V. CHARACTERISTICS OF MS-BIOGRAPHS

In this section, we investigate the characteristics of these graphs and compare them to other real-world graphs.

We offer five views for the data presented in this section:
- The *Frequency* plot that for a value indicated by the x-axis (such as a degree, weight, or component size) shows the number of times that value happens and based on the log-scaled left y-axis,
- The *Fibonacci Binned Frequency* plot based on the log-scaled left y-axis (that connects averaged values of frequency over intervals whose lengths are Fibonacci numbers [38]) to help better visual interpreting of the "cloud of points" that is seen in the tail of frequency plots,
- The *Complementary Cumulative Frequency* plot [39], which is the numerosity-based equivalent of the complementary cumulative distribution function and for a value on the x-axis shows the number of larger or equal values based on the log-scaled left y-axis, and
- The *Cumulative Frequency* plot that for a value on the x-axis shows the number of smaller or equal values as a percentage on the linear-scaled *right* y-axis, and
- The *Cumulative Edges* plot that for a degree indicated by the x-axis shows the total edges of the vertices with degrees less or equal to that degree as a percentage of the total edges and based on the linear-scaled *right* y-axis.

Both in binned plots and complementary cumulative frequency plots in log-log scale data approximately distributed as a power law is displayed on a straight line; they are more reliable than frequency plots for visual inspection [38], [39]. Please note that all functions shown are discrete, and the lines connecting their points are only visual aids.

### A. General Statistics

**Naming.** The name of each graph is started by two characters **M** and **S** as initials of Metaclust (as the source dataset) and Sequence similarity (as the real-world domain of the graph), respectively. The name of the directed subgraphs has a third character **A** that indicates the graph is asymmetric. The name of subgraphs is followed by **3 digits** that show the relative-size of the subgraph in comparison to the MS graph [40], multiplied by a thousand.

Column 5 of Table II summarizes the naming scheme. For the undirected subgraphs MS200 [41], MS50 [42], and MS1 [43] the weight of edges (shown as W in the table) has been considered as the filtering metric. MSA500 [44] is the asymmetric graph of MS. For the directed subgraphs MSA200 [45], MSA50 [46], and MSA10 [47] the vertex-relative weight (shown as VRW in the table) has been used as sampling metric (as explained in Section IV-D, Step 6).

**Statistics.** Table II shows statistics of the MS-BioGraphs: number of vertices and edges, maximum (in-/out-) degree, minimum and maximum values of weights, number of zero (in-/out-) degrees, and average degree. Table II also includes information about the connectivity of MS-BioGraphs: the number of connected components and the relative size of the largest component in comparison to its graph size. We detail the connectivity distributions and its computing process in Section V-D.

The last columns of Table II shows the size of compressed graphs on secondary storage. As we explained in Subsection III-B, a weighted graph is stored as two compressed graphs: the baseline (or underlying) graph (that includes degree of vertices and endpoints of edges) and the labels graph (that contains weights of edges).

### B. Degree Distribution

Figure 3 compares the degree distribution of the MS graph with symmetric versions of Twitter MPI[13][31], [48] (as a social network) and SK-Domain[14] (as a web graph).

The Frequency degree distribution plot shows that the **MS graph has a skewed degree distribution**. The Fibonacci Binned plot shows that the degree distribution does not follow a particular known degree distribution, especially given that two changes of concavity are observed.

---

[13]http://networkrepository.com/soc-twitter-mpi-sws.php
[14]https://law.di.unimi.it/webdata/sk-2005/

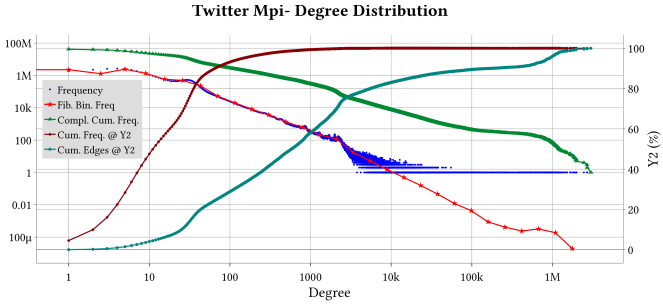

(a) Twitter MPI

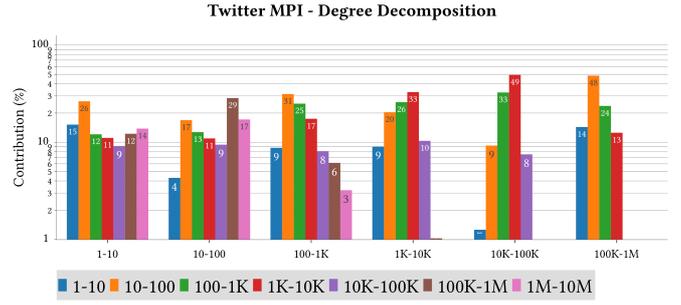

(a) Twitter MPI

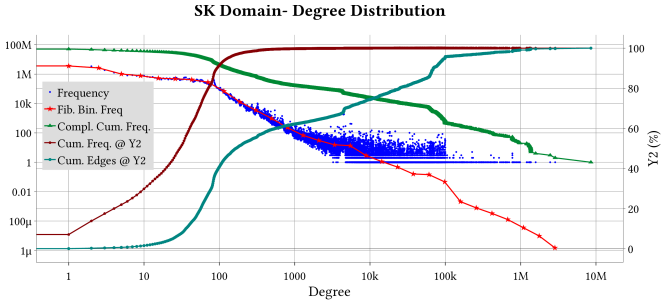

(b) SK-Domain

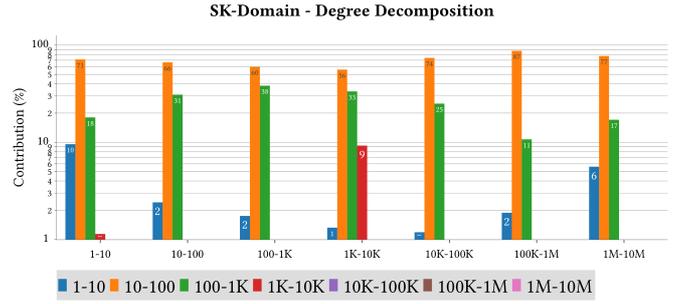

(b) SK-Domain

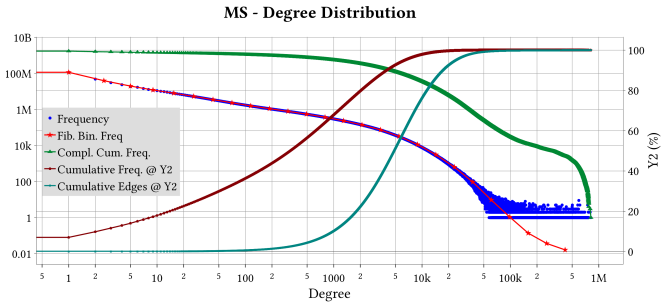

(c) MS

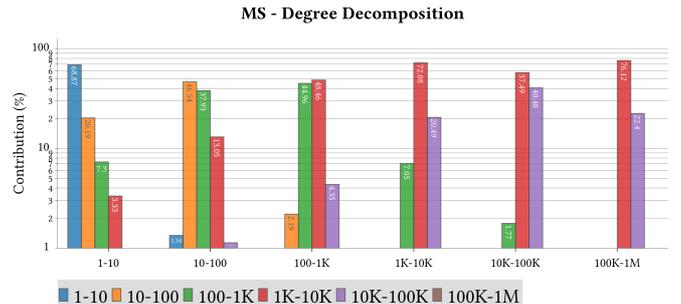

(c) MS

Fig. 3: Degree distribution

Fig. 4: Degree decomposition

By comparing the MS graph to two other types, we identify that the MS graph has a steep slope on the Cumulative Edges plot that indicates more than 98% of edges are incident to the vertices with degrees 100 to 50K. For the Twitter and SK graphs the vertices with degrees between 100–50K containt about 60% and 40% of the total edges, respectively. Unlike the two other types, the low-degree vertices (degrees $\leq 100$) and very high-degree vertices (degrees $\geq 50K$) hardly contribute to the total edges in MS.

To identify the connection between vertices, we use the **degree decomposition** plots [12] in Figure 4. Vertices are divided into vertex classes based on their degrees: vertices with degrees 1-10, 10-100, ... . For each vertex class, we consider edges with destination endpoints in this vertex class. For these edges, we identify and aggregate the vertex classes of the source endpoints. This shows how vertices of different vertex classes contribute (as source of edges) to a vertex class. As an example, in Figure 4a, the first vertex class is 1-10 and has 7 bars. The second bar with yellow color indicates 26% contribution from the vertices with degrees 10-100. In other words, vertices with degrees 10-100 are the source endpoints of 26% of the edges to the vertices with degrees 1-10.

The degree decomposition figures show that, unlike the social network and web graphs, in the MS graph the low-degree vertices (vertices with degree less than 100) are the main constituents of the low-degree vertices and do not contribute to the higher vertex classes. Moreover, MS graph has similarities to the social network graph as high-degree vertices (vertices with degrees in 100-100K classes) are tightly connected to each other.

The tight connection between high-degree vertices in the MS graph becomes more important by comparing the Cumulative Frequency of these vertices in MS graph to the social network in Figure 3 that shows more than 60% of the vertices of the MS graph are vertices with degrees in the range 100–50K (this explains also the steep slope on

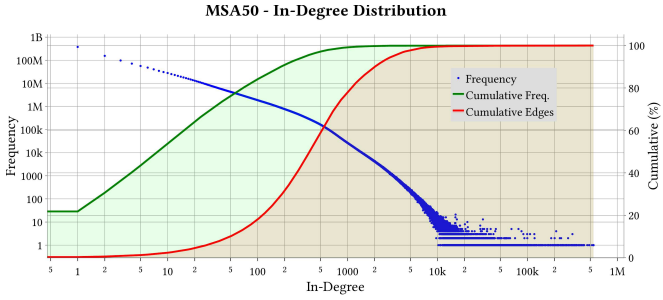

(a) In-degree distribution of MSA50

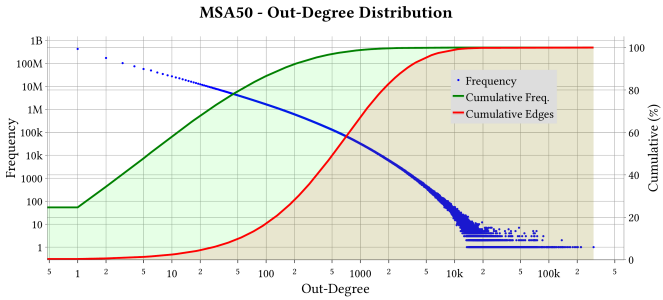

(b) Out-degree distribution of MSA50

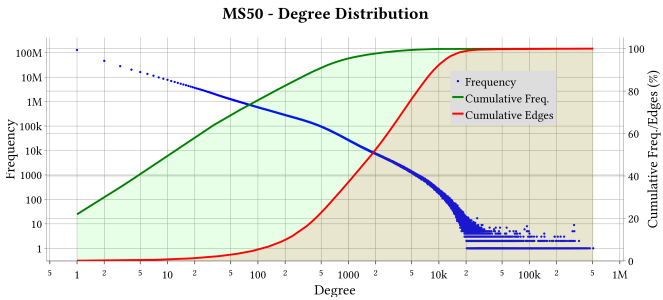

(c) Degree distribution of MS50

Fig. 5: Degree distribution of MSA50 and MS50

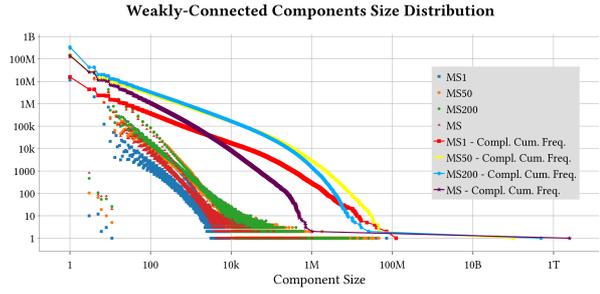

(a) Symmetric graphs

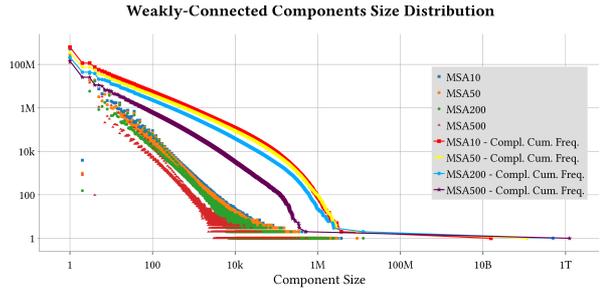

(b) Asymmetric graphs

(c) Weakly-Connected Components size distributions

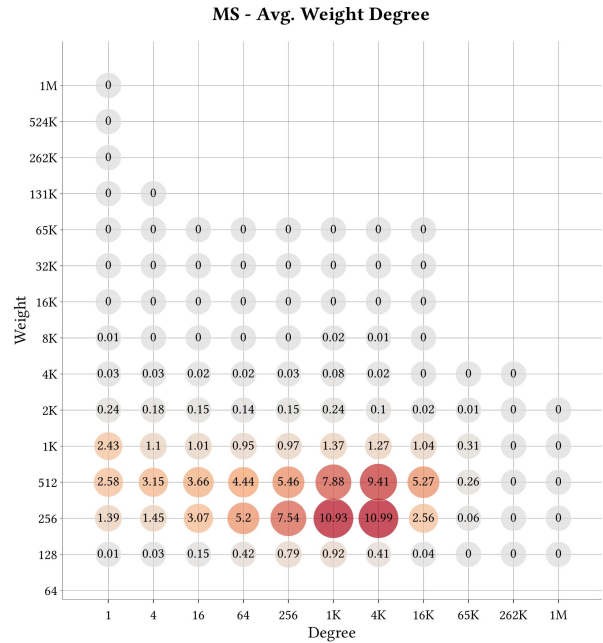

Fig. 7: MS - Avg. weight binned scatter plot

the Cumulative Edges plot). In contrast, these vertex classes include a few percentages of the total vertices in the social network. This **tight connection between high-degree vertices and its coincidence with their high cumulative frequency introduces a new structure of real-world skewed graphs** with obvious differences to the previously studied ones such as web graphs and social networks [12].

We see similar trends in the degree distributions of the MS subgraphs. Figure 5 shows the in-degree distribution of MSA50, the out-degree distribution of MSA50, and the degree distribution of MS50. We see that the slope of the Cumulative Edge reduces and the increase in the curve starts from vertices with lower degrees (degree 2 in MSA50 and degree 10 in MS50 rather than degree 100 in MS), which is a result of filtering methods.

### C. Weight Distribution

Figure 2 shows the weight and vertex-relative weight distribution of the MS graph and their Cumulative Frequency plots. The Figure also includes the Fibonacci Binned plot of weight frequencies. The plots indicate that **weights do not have a random distribution and follow a skewed distribution with a tail close to power-law distribution**.

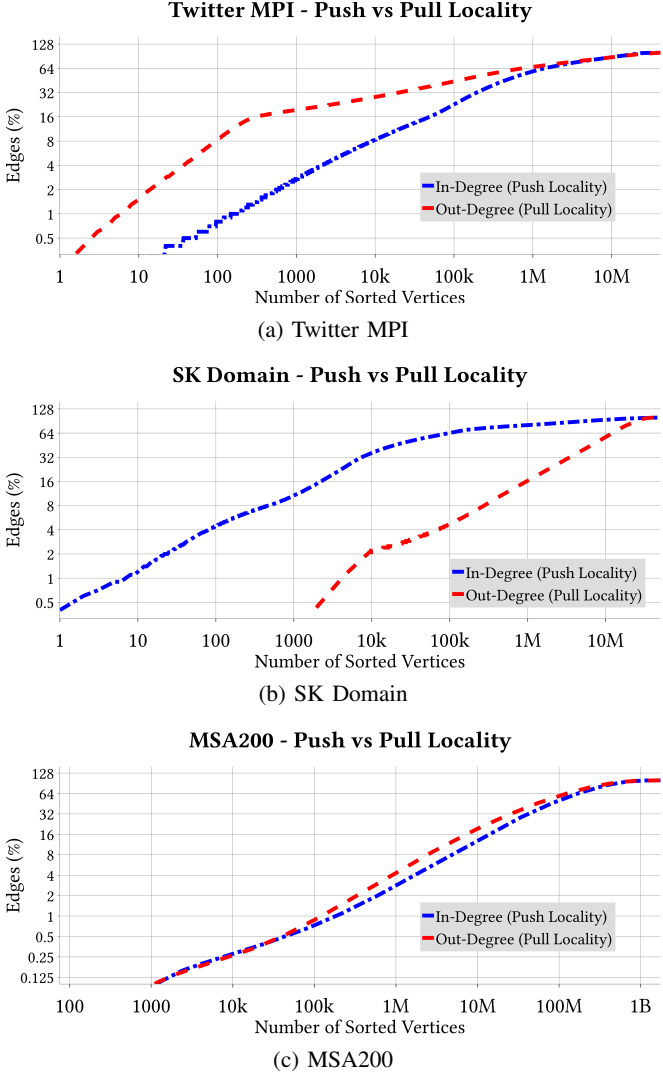

Fig. 8: Push vs Pull Locality

### D. Weakly-Connected Components

Figure 6c shows the component size distribution for symmetric and asymmetric MS-BioGraphs. The plots indicate **a power-law size distribution and a very-high degree of connectivity in MS and also large subgraphs**.

Table II illustrates the number of components in graphs and also the size of the largest component. The table shows that filtering has almost increased the number of components and has reduced the size of the largest component. Moreover, we observe that (as is expected) using vertex-relative weight sampling (in Section IV-D, Step 6) has resulted in better preserving of connectivity in asymmetric subgraphs.

### E. Push vs. Pull Locality

The Push vs. Pull Locality metric [12] considers the cumulative effectiveness of the in-hubs in comparison to the out-hubs in an asymmetric graph. Figure 8 illustrates it for Twitter MPI (as a social network), for SK Domain (as a web graph), and the MSA200 (as a MS-BioGraph).

The push locality curve is created by sorting vertices by their in-degrees in descending order and computing the cumulative number of edges. The x-axis shows the number of sorted vertices and the y-axis shows the cumulative edges (as a percentage of the total edges). The push locality curve illustrates how many of edges are supported by accumulating CPU cache with data of vertices with the largest in-degrees (i.e., in-hubs). Similarly, the pull locality curve is created by using the out-degrees of vertices and illustrates the cumulative edges covered by out-hubs.

Figure 8 shows that for Twitter MPI, the pull locality curve has continuously greater values than the push locality curve. In other words, if we fill cache with the data of out-hubs, more reuse is expected in comparison to filling cache with the data of the same number of in-hubs. On the other hand, for SK Domain, we observe that in-hubs are more powerful than out-hubs and for the same number of hubs, greater number of edges (i.e., more reuse of vertex data) is covered by the out-hubs in comparison to the in-hubs.

For MSA200 (as shown by Figure 8 and also other asymmetric MS-BioGraphs), the push locality curve is very close to the pull locality curve. This shows that **MS-BioGraphs, in contrast to social networks and web graphs, demonstrate the same Push and Pull Locality**.

Table II shows that hubs (including in-hubs and out-hubs) in MS-BioGraphs have a degree fewer than one million, and we explained in Subsection V-B that high-degree vertices are very frequent. We have a gerater number of high-degree vertices with lower contribution per vertex that results in a smoother slope of the push and locality curves (Figure 8) for MSA200 in comparison to Twitter and SK.

## VI. Related Work

*Impacts of Creating Datasets on Progressing Research Fields.* The progress of scientific fields depends on the existence of real-world challenges. To encourage further research in HPC, challenges such as DIMACS[15], and HPEC[16], have been created. Creating updated datasets has the same effect by keeping the research fields motivated and challenging. As an example, image databases such as MOT[17] have been presented in Computer Vision and real-world graph datasets [29], [27], [28], [31] are actively used in HPGP.

*Sequence Alignment Algorithms.* We used the LAST aligner that provides better performance and reliability (Section IV-B). However, the solutions for constructing large graphs apply equally to other aligners [49], [34], [35], [50], [51].

*Generating and Processing Graphs.* Unlike the storage-based processing model [4], [3], [52], the distributed-storage processing model [53] divides the total storage between multiple machines that makes the machines dependent on each other for accessing the storage. The progress of parallel and distributed file systems has provided larger bandwidth that requires new processing models. As explained in Section II-C, creating and analyzing graphs deals with graph algorithms such as graph transposition [54], [55], symmetrization, and sorting [56], [26] that requires further investigations.

---

[15] http://archive.dimacs.rutgers.edu/Challenges/
[16] https://www.omgwiki.org/hpec/files/hpec-challenge/
[17] https://motchallenge.net

*Analyzing Graph Structure.* The study of different graph types and their structures has been performed in [29], [57], [58], [59], [12] that present different topological metrics and tools to analyze the differences between different graph types.

## VII. Conclusion & Future Work

To provide a more effective HPGP research environment by accessing realistic and updated datasets with a better coverage of various application-domains, this paper presents solutions for the challenges in creating and compression of large graphs.

We explained the process of creating large graphs as a multi-step HPC process that requires process-wide model-specific engineering and design of data structures and algorithms. We introduced parallel compression in WebGraph framework that facilitates efficient compressing of large graphs.

We demonstrated the effectiveness of our solutions in generating the **MS-BioGraphs**, a family of sequence similarity graphs with up to 2.5 trillion edges which is 6.6 times greater than the previous largest real-world graph. In addition to HPGP benchmarking and networks study, these graphs have several usages in biology.

We presented a comparative study of the characteristics of these graphs that shows a skewed degree distribution and a particular graph structure by exposing a tight connection between frequent high-degree vertices that makes their structure very different from web graphs and social networks.

Further investigations for optimizing the whole process of creating large graphs are necessary. Is it possible to shorten the flow and parallelize/merge the steps and increase reuse of data? Which data structures and algorithms are needed? What impacts are made by different distributed models?

## Acknowledgements

This work was partially supported by (i) the High Performance Computing center of Queen's University Belfast and the Kelvin-2 supercomputer (UKRI EPSRC grant EP/T022175/1) and (ii) the SERICS project (PE00000014) under the NRRP MUR program funded by the EU - NGEU. First author was also supported by a scholarship from the Department for the Economy, Northern Ireland and Queen's University Belfast.